# The Hubble tension: Change in dark energy or a case for modified gravity?

C Sivaram[1], Kenath Arun*[2] and Louise Rebecca[2,3]

[1]Indian Institute of Astrophysics, Bangalore - 560 034, India

[2]Department of Physics and Electronics, CHRIST (Deemed to be University), Bengaluru - 560029, India

[3]Christ Junior College, Bangalore - 560 029, India

**Abstract:** Recently much controversy has been raised about the cosmological conundrum involving the discrepancy in the value of the Hubble constant as implied by Planck satellite observations of the CMBR in the early Universe and that deduced from other distance indicators (for instance using standard candles like supernovae, tip of the Red Giant branch, etc.) in the present epoch. The Planck estimate is about $67 \, kms^{-1}Mpc^{-1}$, while that deduced from distance indicators at the present epoch is around $73 - 74 \, kms^{-1}Mpc^{-1}$. Also the independent determination of the local value of the Hubble constant based on a calibration of the Tip of the Red Giant Branch (TRGB) and applied to Type Ia supernovae found a value of $69.8 \, kms^{-1}Mpc^{-1}$. Here we propose a modification of the gravitational field on large scales as an alternate explanation for this discrepancy in the value of the Hubble constant as implied in the above-mentioned method, i.e., by Planck observations of the CMBR in the early Universe and that deduced from other distance indicators in the present epoch.

*Corresponding author:

e-mail: kenath.arun@christuniversity.in

Orcid ID: 0000-0002-2183-9425

# 1. Introduction

In recent years, with the advent of new technology, the accuracy in measurement of the Hubble constant has improved tremendously. But this has led to a recent tension that could indicate either the need for new physics or for as-yet unaccounted for uncertainties in the measurements. One method of estimating the Hubble constant 'locally' is by calibrating the luminosities of objects, considered as standard candles like Cepheid variables and thereby estimating distances to nearby galaxies. For measurements of distance scales extending beyond the smooth Hubble flow, calibration of the luminosities of objects like Type Ia supernovae are used. Another way to estimate the Hubble constant involves using and extrapolating data from the cosmic microwave background in the early Universe. The two techniques do produce similar values for the Hubble constant, but the values differ significantly [1-7].

There has been progress in reducing known systematic errors in measurements dealing with the increase in the samples of Type Ia supernovae [8-11]. The Wide Field Camera 3 on the Hubble Space Telescope was used to reduce the uncertainty in the local value of the Hubble constant from 3.3% to 2.4%. Most of this improvement comes from recent near-infrared observations of Cepheid variables in eleven galaxies hosting recent type Ia supernovae. This observation more than doubled the sample of reliable type Ia supernovae (i.e., also having a Cepheid-calibrated distance) to a total of 19. From this, the best estimate of Hubble constant is found to be $73.24 \pm 1.74 \, kms^{-1}Mpc^{-1}$ [8]. The Planck satellite has acquired the highest-resolution maps to date of the sky in microwaves, providing the snapshot of the early Universe about 380,000 years after the Big Bang. An analysis of variations in the temperature and polarization maps leads to an excellent agreement with the current standard model of cosmology [12-14]. Fitting the angular power spectrum of fluctuations in the Planck data to the ΛCDM model leads to a Hubble constant value of $67.8 \pm 0.9 \, kms^{-1}Mpc^{-1}$. The Planck value agrees well with the value of $67.3 \pm 1.1 \, kms^{-1}Mpc^{-1}$ obtained from measurements of baryon acoustic oscillations in combination with SNe Ia [15, 16].

New results [17] from the late Universe, reinforced the Hubble value $H_0$ obtained from the results of Supernova for the Equation of State (SH0ES) [8, 18]. Another independent measurement pertaining to the late Universe based on a calibration of the Tip of the Red Giant Branch when applied to Type Ia supernovae gave a Hubble constant value of $69.8 \pm 0.8 \, kms^{-1}Mpc^{-1}$, midway between the values from Planck and SH0ES [19]. This new standard candle uses red giant stars. Due to their location in a galaxy's uncrowded outskirts, red giants are less prone to the contaminating effects of interstellar dust and nearby stars, than Cepheids. Hence it is much easier to measure accurate luminosities for these stars.

Recently de Jaeger et al. [20] used the SNe II as standard candles to obtain an independent measurement of the Hubble constant. Using seven SNe II with host-galaxy distances measured from Cepheid variables or the tip of the red giant branch, they arrive at $H_0 = 75.8^{+5.2}_{-4.9} kms^{-1} Mpc^{-1}$. This estimate favours those obtained from the conventional distance ladder (like Cepheids taken with SNe Ia) and shows a difference of $8.4\ kms^{-1} Mpc^{-1}$ from the Planck + ($\Lambda$CDM) value. The Hubble constant measures the rate of expansion of the Universe so that the above values suggest that the Universe is expanding faster at present than at the time of the CMBR epoch.

It is implied by Reiss and others [21-24] that this is a real discrepancy and not due to observational uncertainties. The difference amounts to about ten percent in the value of the expansion rate, well above the errors and uncertainties.

## 2 Possible solutions to the discrepancy

Several suggestions have been made as to the cause of such a discrepancy. Perhaps a new kind of matter or unknown particle has now become more dominant, adding to the 'push' in the expansion rate. For instance, a recent suggestion [25] involves a calculation of the amount of change in the quantum fields, needed to account for the dark energy (DE) change. This quantum field causing a change in the DE implies the existence of a new particle with a mass roughly that of the axion (about $10^{-3} eV$), already predicted earlier. As the background density decreases this particle is now more dominant.

In general, we can write for a change in Hubble constant ($dH$), relating it to a change in density $d\rho$ as:

$$2HdH = \frac{8\pi G}{3} d\rho \qquad (1)$$

Here, with the values, $H = 70 kms^{-1} Mpc^{-1}$, $dH = 5 kms^{-1} Mpc^{-1}$. This gives, $2HdH = 10^{-36} cm^2/s^2/cm^2$, which from equation (1) gives:

$$d\rho \approx 2 \times 10^{-30} g/cc \qquad (2)$$

This is the required change in density that gives the change in Hubble constant. The quantum field energy density is given as $\frac{m}{(\hbar/mc)^3}$, i.e., a pair of particles separated by Compton length, where the particle mass causing the change in the quantum field energy density is $m$.

Hence, we have:

$$d\rho = \frac{m^4 c^3}{\hbar^3} \tag{3}$$

Equations (1) – (3) imply:

$$m \approx 10^{-36} g = 10^{-4} eV \tag{4}$$

Equation (1) also implies that a change in $H$ could be due to a change in the gravitational constant, rather than a change in density. This would require an increase in $G$ by about 10 per cent over a few billion years. This would contradict solar system measurements (i.e., from radar ranging of interplanetary spacecraft putting limits on varying $G$ theories [26]). However, if not connected with dark energy, change in $G$ can still affect $H$ as in equation (1), if there is just matter density dominating at higher $z$. This would however require increase in $G$ rather than a decrease as in the usual theory. A decrease of $G$ with epoch could increase the DE density since:

$$\rho_\Lambda = \frac{\Lambda c^2}{8\pi G} \tag{5}$$

However as on the RHS of equation (1) we have $G$ multiplying $\rho$, this would cancel out the $G$ variation, i.e., $G\rho_\Lambda = constant$.

## 3 Modification to Newtonian Gravity

We point out that a form of modified gravity – used in recent work [27-29] as a way to account for flat rotation curves of galaxies without invoking DM – could in principle account also for the increase in $H$. This also accounts for the DM needed in galaxy clusters [28, 29] and is consistent with other observations of galaxies in the early Universe (which are more consistent with MOND [30, 31]). In this work, we invoked a minimal acceleration in the gravitational field, $a_0$, which implies a modification of gravity over a scale larger than $R_c$, given by:

$$R_c = \left(\frac{GM}{a_0}\right)^{1/2} \tag{6}$$

$M$ is the mass of the system, either a galaxy or galaxy cluster. Again, the gravitational self-energy density given by $(\nabla\phi)^2$ becomes important if $\rho$ drops below certain value. So that the modified Poisson equation is:

$$\nabla^2\phi + (\nabla\phi)^2 = 4\pi G\rho \tag{7}$$

For the self-energy term to become important the equation $\nabla^2\phi + (\nabla\phi)^2 = 0$, has the solution [32]:

$$\phi = \frac{GM}{r} + K' \ln\frac{R}{R_C} \tag{8}$$

The extra term accounts for flat rotation curves and also for the high outer rotation velocities of super spiral galaxies. The usual Friedmann equation (for a flat Universe) now gets modified to:

$$\frac{\dot{R}^2}{R^2} = \frac{8\pi G\rho}{3} + \frac{(GMa_0)^{1/2}}{R^2}\ln\frac{R}{R_C} + \frac{\Lambda c^2}{3} \qquad (9)$$

Here again, $R_C$ is the scale factor at which the modification becomes relevant. The usual $\Lambda$ term remains unchanged. The second term can be seen as a modification in potential energy due to gravitational self-energy density in the usual balance between kinetic energy term and potential energy term, $\frac{8\pi G\rho}{3}$ (in the expanding Universe), i.e., the usual Newtonian analogue agreeing with the GR result [28, 29].

With $a_0 = 10^{-8} cm/s^2$ [33, 34], and considering the Universe as having expanded at present to $R > R_{min}$ (where $R_{min} \approx 10^{28} cm$) the modified term will also contribute. Putting in the appropriate values, i.e., $R = 2 \times 10^{28} cm, \rho = 10^{-29} g/cc$, and the mass of the Universe, $M = 2\pi^2 R^3 \rho \approx 10^{56} g$, the usual first term is $\approx 10^{-35}$, whereas the second term $\approx 10^{-36}$ which now gives for Hubble constant $H_0 \approx 70 \, kms^{-1}Mpc^{-1}$ (a difference of about five per cent).

This would suggest that this extra term now manifesting itself would cause an increase of the expansion rate, i.e., a change in Hubble constant by about five per cent. This could perhaps account for the faster expansion rate seen at the present epoch. In the early Universe, the second term was much smaller as it is proportional to $\ln r$. At present this term is comparable with the first term.

The extra term in the Poisson equation (equation (7)) and its solution given by equation (8) could also have interesting consequences for current observations of super spirals [35, 36], wherein their large extent (450,000 light-years) is associated with large rotation velocities of up to $\sim 450 km/s$ at their periphery. The rotation velocity on the outskirts of the super spirals follows from the usual Newtonian gravity as,

$$v_0 = \sqrt{\frac{GM_T}{R}} \qquad (10)$$

where $M_T$ is the total mass of the galaxy, including DM. Conventionally such a large velocity would imply a colossal amount of DM, i.e., about $\sim 10^{13} M_\odot$. However, the extra term in our model would give a velocity:

$$v = (GMa_0)^{\frac{1}{4}}\left(\ln\frac{R}{R_C}\right)^{1/2} \qquad (11)$$

where $R_C$ corresponds to the radius at which acceleration approaches $a_0$. With $R_C = 20 kpc$ and super spiral extant $R = R_{SS} \approx 200 kpc$, this would give velocities $\sim 450 km/s$. I.e., the logarithmic term makes gravity stronger above $R_C$ (i.e., potential going as $\ln r$, instead of $1/r$), so that we do not need such colossal amounts of DM. This is in good agreement with observations as shown in table 1.

**Table 1** Velocities of super spiral galaxies at the periphery with DM ($v_0$) and with modified gravity ($v$)

| Name | $r$ (kpc) | $M_T$ ($M_\odot$) | $M$ ($M_\odot$) | $v_o$ ($km\ s^{-1}$) | $v$ ($km\ s^{-1}$) |
|---|---|---|---|---|---|
| 2MASX J0304263+0418219 | 30 | $10^{12}$ | $5 \times 10^{11}$ | 342 | 399 |
| 2MASX J10222648+0911396 | 33 | $10^{12}$ | $6.7 \times 10^{11}$ | 311 | 333 |
| 2MASX J11483552+0325268 | 31 | $10^{12}$ | $2.9 \times 10^{11}$ | 324 | 338 |
| 2MASX J12422564+0056492 | 14 | $4. \times 10^{11}$ | $1.8 \times 10^{11}$ | 289 | 279 |
| 2MASX J12592630−0146580 | 20 | $8 \times 10^{11}$ | $1.8 \times 10^{11}$ | 318 | 334 |
| 2MASX J13033075−0214004 | 23 | $8 \times 10^{11}$ | $2.5 \times 10^{11}$ | 308 | 350 |
| SDSS J143447.86+020228.6 | 26 | $10^{12}$ | $4. \times 10^{11}$ | 344 | 390 |
| 2MASX J15404057−0009331 | 30 | $10^{12}$ | $2.7 \times 10^{11}$ | 304 | 334 |
| 2MASX J16184003+0034367 | 40 | $1.5 \times 10^{12}$ | $5 \times 10^{11}$ | 384 | 360 |
| 2MASX J20541957−0055204 | 37 | $10^{12}$ | $3 \times 10^{11}$ | 317 | 322 |
| 2MASX J21362206+0056519 | 29 | $10^{12}$ | $3 \times 10^{11}$ | 336 | 347 |
| 2MASX J21431882−0820164 | 18 | $6 \times 10^{11}$ | $1.4 \times 10^{11}$ | 317 | 319 |
| 2MASX J22073122−0729223 | 31 | $6 \times 10^{11}$ | $4. \times 10^{11}$ | 243 | 280 |
| 2MASX J23130513−0033477 | 19 | $6 \times 10^{11}$ | $4. \times 10^{11}$ | 292 | 331 |

$v_o$ is the maximum speed at radius $r\ (kpc)$ from observation and $v$ is that estimated from equation (11).

**4 Conclusion**

One of the most intriguing puzzles in cosmology today is that of the Hubble tension involving the measurement of Hubble constant $H_0$, which gives the rate of expansion of the Universe. As noted above, the measurement of $H_0$ using the Planck satellite study of the cosmic microwave background radiation and the value of $H_0$ obtained from observing stars and galaxies in the late Universe, do not coincide. This discrepancy has been ongoing for years, increasing as each new study – of both the early and late Universe – yield ever more precise results. This Hubble tension could be either due to error in measurements (in the early or the late Universe observations) or may be an actual indication of the present expansion rate of the Universe.

In this work, we have proposed a possible solution to this ongoing Hubble tension involving the discrepancy in the value of the Hubble constant obtained from different methods. Our model invokes a minimal acceleration ($a_0$) in the gravitational field which implies a modification of gravity over a scale larger than a critical size, $R_c$. This model would imply that a particular modification of the gravitational field on large scales (low accelerations) could well provide a possible solution for this discrepancy and perhaps help in relaxing the Hubble tension.

This modification of gravity was used earlier to account for rotation curves of galaxies and for cluster dynamics without invoking dark matter. When this model was applied to the recent observations of super spiral galaxies it reduces the need for copious amounts of dark matter, which would be required otherwise. We also predict that as this extra term increases slowly with the logarithm of the scale factor, this would give rise to an even higher value of $H$ at later epochs.